	\newcommand{\et}{\hspace{-0.08in}{\bf .}\hspace{0.1in}}
     
 	\newcommand{\BOX}{\hbox {$\sqcap$ \kern -1em $\sqcup$}}
	\newcommand{\qed}{\hskip 3em \hbox{\BOX} \vskip 2ex}
	
	\newcommand{\Inv}{{\rm Inv}}
	\newcommand{\tensor}{\otimes}
	\newcommand\C{{\bf C}}
	\def\section#1{\vskip3em{\centerline {\bf#1}}\vskip3em}
	\newcommand\R{{\bf R}}
	\newcommand{\be}{\begin{equation}}
     \newcommand{\ee}{\end{equation}}
     \newcommand{\ba}{\begin{eqnarray}}
     \newcommand{\ea}{\end{eqnarray}}
     \newcommand{\ban}{\begin{eqnarray*}}
     \newcommand{\ean}{\end{eqnarray*}}
	\newcommand{\A}{{\cal A}}
	\newcommand{\G}{{\cal G}}
     \renewcommand{\c}{{\cal C}}
     \newcommand{\maps}{\colon}

	\newcommand{\Diff}{{\rm Diff}}

	\newcommand{\Fun}{{\rm Fun}}
	\newcommand{\Ad}{{\rm Ad}}

	\newtheorem{theorem}{Theorem}
	
	\newtheorem{lemma}{Lemma}

	\newcommand{\SL}{{\rm SL}}
	\newcommand{\SU}{{\rm SU}}

	\documentstyle[12pt]{article}
\begin{document}

	\begin{center}
	{\bf  Spin Network States in Gauge Theory \\}
	\vspace{0.5cm}
	{\em John C. Baez\\}
	\vspace{0.3cm}
	{\small Department of Mathematics \\
	University of California\\
        Riverside CA 92521\\}
	\vspace{0.3cm}
	{\small November 1, 1994}
	\vspace{0.3cm}
	\end{center}

\begin{abstract} Given a real-analytic manifold $M$, a compact connected
Lie group $G$ and a principal $G$-bundle $P \to M$, there is a canonical
`generalized measure' on the space $\A/\G$ of smooth connections on $P$
modulo gauge transformations.  This allows one to define a Hilbert space
$L^2(\A/\G)$.  Here we construct a set of vectors spanning $L^2(\A/\G)$.
These vectors are described in terms of `spin networks': graphs $\phi$
embedded in $M$, with oriented edges labelled by irreducible unitary
representations of $G$, and with vertices labelled by intertwining
operators from the tensor product of representations labelling the
incoming edges to the tensor product of representations labelling the
outgoing edges.  We also describe an orthonormal basis of spin networks
associated to any fixed graph $\phi$.  We conclude with a discussion of
spin networks in the loop representation of quantum gravity, and give a
category-theoretic interpretation of the spin network states.
\end{abstract}

\section{Introduction}

Penrose \cite{Penrose} introduced the notion of a spin network as an
attempt to go beyond the concept of a manifold towards a more
combinatorial approach to spacetime.  In his definition, a spin network
is a trivalent graph labelled by spins $j = 0, {1\over 2}, 1, \dots, $
satisfying the rule that if edges labelled by spins $j_1, j_2, j_3$ meet
at a vertex, the Clebsch-Gordon condition holds:
\[ |j_1 - j_2| \le j_3 \le j_1 + j_2. \]
In fact, the spins should be thought of as
labelling finite-dimensional irreducible representations of $\SL(2,\C)$,
and the Clebsch-Gordon condition is necessary and sufficient for there
to be a nontrivial intertwining operator from $j_1 \tensor j_2$ to $j_3$.
One can use the representation theory of $\SL(2,\C)$ to obtain
numerical invariants of such labelled graphs.

Recently, spin networks and their generalizations have played an
important role in topological quantum field theories such as
Chern-Simons theory \cite{RT} and the Turaev-Viro model \cite{KL,TV} in
dimension three, and the Crane-Yetter model \cite{CKY} in dimension
four.  In these theories, the category of finite-dimensional
representations of $\SL(2,\C)$ is
replaced by a suitable category of representations of a quantum group.
Again, the key idea is a method for obtaining invariants of graphs whose
edges are labelled by irreducible representations.  Here, however, the
graphs are regarded as embedded in $\R^3$, and are equipped with a
framing.  The edges are oriented, and reversing the orientation of an
edge has the same effect as replacing the representation $\rho$
labelling it by the dual representation $\rho^\ast$.  Moreover, the
graphs need not be trivalent, but each vertex must be labelled with an
intertwining operator  from the tensor product of the representations
labelling the incoming edges to the tensor product of the
representations labelling the outgoing edges.  The $\SL_q(2)$ case is
very similar to the situation studied by Penrose, and effectively
reduces to it in the limit $q \to 1$.  In this case every representation
is self-dual, so edges do not need orientations, and the space of
intertwining operators from $j_1 \tensor j_2$ to $j_3$ is at most
1-dimensional, so trivalent vertices do not need labels.

In parallel with these developments, mathematical work on the loop
representation of quantum gravity \cite{Ashtekar} has led to a theory of
functional integration on spaces $\A/\G$ of connections modulo gauge
transformations which allows one to define a rigorous version of the
Hilbert space $L^2(\A/\G)$.  This space is not defined using the purely
formal `Lebesgue measure' on $\A/\G$, but instead using the canonical
`generalized measure' on $\A/\G$ coming from Haar measure on the
(compact, connected) gauge group $G$.  Here we construct an explicit
set of vectors spanning $L^2(\A/\G)$ using spin networks.  Each of these
`spin network states' $\Psi_{\phi,\rho,\iota}$ is labelled by a choice
of: a) an oriented, unframed graph embedded in the base manifold $M$, b)
a labelling of each edge $e$ of $\phi$ by an irreducible representation
$\rho_e$ of $G$, and c) a labelling of each vertex $v$ of $\phi$ by a
vector $\iota_v$ in the space of intertwining operators from
the tensor product of `incoming' representations to the tensor product
of the `outgoing' representations.

In the language of canonical quantum gravity, vectors in
$L^2(\A/\G)$ represent states at the `kinematical' level.  Spin
networks have also been used by the physicists Rovelli and Smolin
to describe states at the `diffeomorphism-invariant' level
\cite{RS2}. Mathematically, states at the diffeomorphism-invariant
level are thought to be given by diffeomorphism-invariant
generalized measures on $\A/\G$. Such states have already been
characterized, and examples constructed, using the language of
graphs \cite{Baez}.  To give a rigorous foundation to the work of
Rovelli and Smolin, one would like to construct large classes of
such states using spin networks.  This is likely to involve a
generalization of techniques due to Ashtekar {\it et al }
\cite{Ashtekar3,AL2} for constructing diffeomorphism-invariant
generalized measures from certain knot invariants.  The present
work is intended as a first step in this direction.  For
more remarks on the applications of spin network states
to quantum gravity, as well as a category-theoretic
interpretation of the spin network states, see Section 5.

\section{Gauge Theory on a Graph}

In this section we develop the basic concepts of gauge theory on a
graph, which we apply in the next section to gauge theory on manifolds.
Readers familiar with lattice gauge theory may find it useful to think
of what follows as a slight generalization of gauge theory on a finite
lattice.  In the case of a trivial bundle, a connection on a graph will
assign a group element to each edge of the lattice, or `bond', and a
gauge transformation will assign a group element to each vertex, or
`site'.  In fact, we only consider trivializable bundles.  However, to
apply our results to gauge theory on manifolds, it is convenient not to
assume the bundles are equipped with a fixed trivialization.

Let $G$ be a compact Lie group, and let $\phi$ be a
(finite, directed) {\it graph}, by which we mean a finite set $E$ of {\it
edges}, a finite set $V$ of {\it vertices}, and functions
\[   s\maps E \to V, \qquad t \maps E \to V
.\]
We call the vertex $s(e)$ the {\it source} of the edge $e$, and the
vertex $t(e)$ the {\it target} of $e$.  Let $P$ be a principal
$G$-bundle over $V$, regarding $V$ as a space with the discrete
topology.  Any such bundle is trivializable, but we do not assume $P$ is
equipped with a fixed trivialization.  Given any vertex $v$, we write
$P_v$ for the fiber of $P$ over $v$.

Given any edge $e$, let $\A_e$ denote the space of smooth maps
$F\maps P_{s(e)} \to P_{t(e)}$ that are compatible with the right action
of $G$ on $P$:
\[            F(xg) = F(x)g .\]
We define $\A$, the space of {\it connections} on $\phi$, by
\[      \A = \prod_{e \in E} \A_e .\]
Given $A \in \A$, we write $A_e$ for the value of $A$ at the edge
$e \in E$.  Note a trivialization of $P$ lets us to identify each
space $\A_e$ with a copy of $G$, with elements acting as maps from
$P_{s(e)} \cong G$ to $P_{t(e)} \cong G$ by left multiplication.  We may
then identify $\A$ with $G^E$.  This equips $\A$ with the
structure of a smooth manifold, and also endows it with a probability
measure $\mu$, namely the product of copies of normalized Haar
measure on $G$.  One can check that this smooth structure and measure
are independent of the choice of trivialization.  It follows that the
space $C(\A)$ of continuous functions on $\A$ and the Hilbert
space $L^2(\A)$ of square-integrable functions on $\A$ are
well-defined in a trivialization-independent manner.

Similarly, we define the group of gauge transformations on $\phi$, written
$\G$, by
\[       \G = \prod_{v \in V} P_v \times_{\Ad} G  .\]
This is just the usual group of gauge transformations of the bundle
$P$, so a trivialization of $P$ allows us to identify $\G$
with $G^V$.
Given $g \in \G$, we write $g_v$ for the value of $g$ at $v$.
We may regard $g_v$ as a map from $P_v$ to itself, so that
the group $\G$ acts on $\A$ by
\[       (gA)_e = g_{t(e)}\, A_e\, g_{s(e)}^{-1}  .\]
Note that $\G$ acts on $\A$ in a smooth, measure-preserving
manner.

We can push forward the measure on $\A$ to a probability measure on
$\A/\G$ using the quotient map $\A \to \A/\G$.  The space $L^2(\A/\G)$
is then naturally isomorphic to the $\G$-invariant subspace of
$L^2(\A)$.  In what follows, we describe an orthonormal basis of
$L^2(\A/\G)$ using spin networks.

Let $\Lambda$ denote a set of irreducible unitary
representations of $G$, one for each equivalence class.
We assume the trivial representation of $G$ on
$\C$ is a member of $\Lambda$.
Given $\rho \in \Lambda$, we write $\rho^\ast$ for the
representation in $\Lambda$ equivalent to the dual of $\rho$.
Recall that $G \times G$ acts on $G$ by
\[    (g_1,g_2)(g) = g_2gg_1^{-1} ,\]
and that this makes $L^2(G)$ into a unitary representation of $G\times
G$, which by the Peter-Weyl theorem is isomorphic to
\[   \bigoplus_{\rho \in \Lambda} \rho \tensor \rho^\ast  .\]

As an immediate consequence we have:
\begin{lemma} \label{lem1} \et Any
trivialization of $P$ determines a unitary equivalence of the
following representations of $\G$:
\[     L^2(\A) \cong \bigotimes_{e \in E}
\bigoplus_{\rho \in \Lambda} \rho \tensor \rho^\ast, \]
where $g \in \G$ acts on the latter space by
\[        \bigotimes_{e \in E}
\bigoplus_{\rho \in \Lambda} \rho(g_{s(e)}) \tensor
\rho^\ast(g_{t(e)})  .\]
\end{lemma}

However, to describe the spin network states, a slightly
different description of $L^2(\A)$ is preferable.  An element
$\rho \in \Lambda^E$ is a labelling of all edges $e \in E$ by
irreducible representations $\rho_e \in \Lambda$.  In
these terms, when we fix a trivialization of $P$ we obtain
a unitary equivalence
\[       L^2(\A) \cong \bigoplus_{\rho \in \Lambda^{\! E}} \;
\bigotimes_{e \in E} \rho_e \tensor \rho_e^\ast ,\]
with $g \in \G$ acting on the right hand side by
\[  \bigoplus_{\rho \in \Lambda^{\! E}}\;
\bigotimes_{e \in E} \rho_e(g_{s(e)}) \tensor
\rho_e^\ast(g_{t(e)}) .\]

Now, given a vertex $v \in V$, let $S(v)$ denote the set of all edges of
$\phi$ having $v$ as source, and $T(v)$ the set of all edges having $v$
as target.  Then the above formula for $L^2(\A)$ gives:
\begin{lemma}\label{lem2}\et   Any trivialization of $P$
determines a unitary equivalence of the following representations of
$\G$:
\[     L^2(\A) \cong \bigoplus_{\rho \in \Lambda^{\! E}}\;
\bigotimes_{v \in V} \left( \bigotimes_{e \in S(v)}
\rho_e \, \otimes \ \bigotimes_{e \in T(v)}
\rho_e^\ast \right) ,\]
where $g \in \G$ acts on the latter space by
\[ \bigoplus_{\rho\in \Lambda^{\! E}} \;
\bigotimes_{v \in V} \left( \bigotimes_{e \in S(v)}
\rho_e(g_v) \, \otimes  \bigotimes_{e \in T(v)}
\rho_e^\ast(g_v) \right)   .\]
\end{lemma}

This allows us to describe $L^2(\A/\G)$ in terms of spin networks
as follows.   For each vertex
$v$ of the graph $\phi$, and for each choice $\rho \in \Lambda^E$
of labellings of the edges of $\phi$ by irreducible
representations of $G$, let $\Inv(v,\rho)$ denote the subspace of
invariant elements of the following representation of $G$:
\[  \bigotimes_{e \in S(v)}
\rho_e  \otimes  \bigotimes_{e \in T(v)}
\rho_e^\ast  \]
Note that elements $f \in \Inv(v,\rho)$
may be thought of as intertwining operators
\[       f \maps \bigotimes_{e \in S(v)} \rho_e
\to \bigotimes_{e \in T(v)} \rho_e .\]
Lemma \ref{lem2} implies the following:
\begin{lemma}\label{lem2.5}\et
In a manner independent of the choice of trivialization of
$P$,  $L^2(\A/\G)$ is isomorphic as a Hilbert space to
\[  \bigoplus_{\rho\in \Lambda^{\! E}} \bigotimes_{v \in V} \Inv(v,\rho)
.\]
\end{lemma}
As a consequence, $L^2(\A/\G)$ is spanned by {\it spin network states}
\[ \Psi_{\rho,\iota} = \bigotimes_{v \in V} \iota_v, \]
where $\rho \in
\Lambda^E$ is any labelling of the edges $e$ of $\phi$ by irreducible
representations $\rho_e \in \Lambda$, and $\iota_v \in \Inv(v,\rho)$ for
each vertex $v$ of $\phi$.  In particular, if we let
$\rho$ range over $\Lambda^E$ and for each vertex $v$ let
$\iota_v$ range over an orthonormal basis of $\Inv(v,\rho)$, the spin
network states $\Psi_{\rho,\iota}$ form an orthonormal basis of
$L^2(\A/\G)$.

To deal with gauge theory on manifolds we also need to study the
dependence of $L^2(\A/\G)$ on the graph $\phi$, and particularly
the situation where a graph $\psi$ is included in the graph $\phi$.
For this we need to write subscripts such as $\psi$ or
$\phi$ on the symbols $E$, $V$, $P$, $\A$, $\G$, $\mu$, $S(v)$,
$T(v)$ and $\Inv(v,\rho)$ to indicate the dependence on the
graph.

Here is an example of what we have in mind by a
graph $\psi$ being `included' in a graph $\phi$.

\begin{center}
\setlength{\unitlength}{0.0125in}%
\begin{picture}(105,160)(335,360)
\thicklines

\put(270,440){\oval( 80, 80)[br]}
\put(270,440){\oval( 80, 80)[bl]}
\put(270,400){\circle*{5}}
\put(275,390){\makebox(0,0)[lb]{\raisebox{0pt}[0pt][0pt]{$v_4$}}}
\put(270,520){\circle*{5}}
\put(275,520){\makebox(0,0)[lb]{\raisebox{0pt}[0pt][0pt]{$v_1$}}}
\put(270,360){\circle*{5}}
\put(275,360){\makebox(0,0)[lb]{\raisebox{0pt}[0pt][0pt]{$v_5$}}}
\put(215,440){\makebox(0,0)[lb]{\raisebox{0pt}[0pt][0pt]{$v_2$}}}
\put(230,440){\circle*{5}}
\put(230,430){\vector(0,-1){10}}
\put(310,440){\circle*{5}}
\put(315,440){\makebox(0,0)[lb]{\raisebox{0pt}[0pt][0pt]{$v_3$}}}

\put(310,430){\vector(0,-1){10}}
\put(270,480){\line( 0,-1){ 80}}
\put(270,475){\vector( 0,-1){ 10}}
\put(270,400){\makebox(0.4444,0.6667){\tenrm .}}
\put(270,520){\line( 0,-1){ 40}}
\put(270,400){\line( 0,-1){ 40}}
\put(270,385){\vector( 0,-1){ 10}}

\put(365,440){\makebox(0,0)[lb]{\raisebox{0pt}[0pt][0pt]{$\hookrightarrow$}}}

\put(470,440){\oval( 80, 80)[br]}
\put(470,440){\oval( 80, 80)[tr]}
\put(470,440){\oval( 80, 80)[tl]}
\put(470,440){\oval( 80, 80)[bl]}
\put(470,400){\circle*{5}}
\put(475,390){\makebox(0,0)[lb]{\raisebox{0pt}[0pt][0pt]{$v_4$}}}
\put(470,480){\circle*{5}}
\put(475,485){\makebox(0,0)[lb]{\raisebox{0pt}[0pt][0pt]{$v_6$}}}
\put(470,520){\circle*{5}}
\put(475,520){\makebox(0,0)[lb]{\raisebox{0pt}[0pt][0pt]{$v_1$}}}
\put(470,360){\circle*{5}}
\put(480,360){\makebox(0,0)[lb]{\raisebox{0pt}[0pt][0pt]{$v_5$}}}

\put(430,430){\vector(0,-1){10}}
\put(430,440){\circle*{5}}
\put(430,450){\vector(0,1){10}}

\put(415,440){\makebox(0,0)[lb]{\raisebox{0pt}[0pt][0pt]{$v_2$}}}
\put(510,440){\circle*{5}}
\put(515,440){\makebox(0,0)[lb]{\raisebox{0pt}[0pt][0pt]{$v_3$}}}
\put(510,450){\vector(0,1){10}}
\put(510,430){\vector(0,-1){10}}
\put(470,480){\line( 0,-1){ 80}}
\put(470,445){\vector( 0,-1){ 10}}
\put(470,400){\makebox(0.4444,0.6667){\tenrm .}}
\put(470,520){\line( 0,-1){ 40}}
\put(470,400){\line( 0,-1){ 40}}
\put(470,495){\vector( 0,1){ 10}}
\put(470,385){\vector( 0,-1){ 10}}
\end{picture}
\vskip 2em
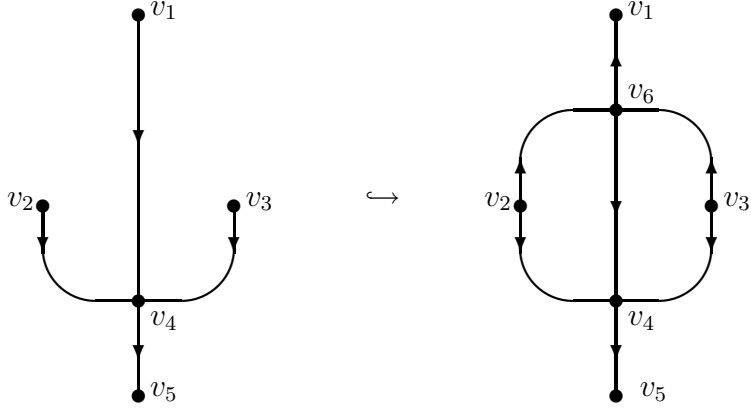
{\tenrm Fig.\ 1.  Graph $\psi$ included in graph $\phi$}
\end{center}

\noindent  Here we have labelled the vertices of $\psi$ and
$\phi$ but not the edges.  Note that every vertex of $\psi$ is a vertex of
$\phi$, but edges of $\psi$ may be `products' of edges
in $\phi$ and their inverses.

To
precisely define the notion of one graph being `included' in
another, we first define a {\it path} in a graph $\phi$ to be a
sequence of vertices $v_1, \dots, v_n \in V_\phi$, together
with, for each $i$, $1 \le i \le n$, an edge $f_i \in E_\phi$ such that
either:
\be       s(f_i) = v_i, \qquad t(f_i) = v_{i+1},  \label{1} \ee
or
\be       t(f_i) = v_i , \qquad s(f_i) = v_{i+1}.  \label{2} \ee
In this situation, we write the path as a product $f_1^{\pm 1} \cdots
f_n^{\pm 1}$, where the exponents are either $+1$ or $-1$ depending on
whether case (\ref{1}) or case (\ref{2}) holds, and we say that the edges
$f_1, \dots, f_n$ {\it appear} in the path.  We say the path is {\it
simple} if the vertices $v_1, \dots, v_n$ are distinct, with the
exception that we allow $v_1 = v_n$.  Heuristically, a simple path in
is one that never retraces or intersects
itself, the only exception being that it may end where it began.

If we have graphs $\psi$ and $\phi$,  an {\it inclusion} of
$\psi$ in $\phi$,  written $i \maps \psi \hookrightarrow \phi$,
is a one-to-one map $i \maps V_\psi \to V_\phi$ together with an
assignment to each edge $e \in E_\psi$ of a simple path $i(e)$ in
$\phi$ from $i(s(e))$ to $i(t(e))$, such that each edge of
$\phi$ appears in at most one path $i(e)$.  In the rest of this section we
assume we are given an inclusion $i\maps \psi \hookrightarrow
\phi$ and principal $G$-bundles $P_\psi \to V_\psi$, $P_\phi \to
V_\phi$ such that $i^\ast P_\phi = P_\psi$.  To simplify the
notation we assume, without loss of generality, that
$V_\psi$ is a subset of $V_\phi$ and $i \maps V_\psi \to V_\phi$
is given by $i(v) = v$.  In this situation the bundle $P_\psi$ is
just the restriction of $P_\phi$ to $V_\psi$.

In this situation there is a map
\[   i^\ast \maps \A_\phi \to \A_\psi  \]
given as follows.  Recall that a connection $A$ on $\phi$
assigns to each edge $f$ of $\phi$ a map $A_f \maps P_{s(f)} \to
P_{t(f)}$ compatible with the right $G$-action on $P_\phi$.
Similarly, the connection $i^\ast(A)$ on $\psi$ must assign to
each edge $e$ of $\psi$ a map $i^\ast(A)_e \maps P_{s(e)} \to
P_{t(e)}$ compatible with the right $G$-action.  If the inclusion
$i$ assigns to $e$ the simple path
\[   i(e) = f_1^{\pm 1} \cdots f_n^{\pm 1} ,\]
we let
\[   i^\ast(A)_e = A_{f_n}^{\pm 1} \cdots A_{f_1}^{\pm 1} .\]
One can check that $i^\ast(A)$ is well-defined and indeed a connection
on $\psi$.  The order-reversal here is due to the unfortunate fact that
the convention for writing products of paths is the opposite of that for
composites of maps.

By arguments already given in the
more concrete cases treated earlier \cite{ALMMT,Baez2,L}, the
map $i^\ast$ is smooth and onto, and the measure $\mu_\phi$ on
$\A_\phi$ pushes forward by $i^\ast$ to give the measure
$\mu_\psi$ on $\A_\psi$.  This implies that $i^\ast$
yields a one-to-one algebra homomorphism from $C(\A_\psi)$ to
$C(\A_\phi)$, which we write simply as
\[  i \maps C(\A_\psi) \to C(\A_\phi) ,\]
and also that $i^\ast$ yields an isometry
\[  i \maps L^2(\A_\psi) \to L^2(\A_\phi) .\]
There is also a surjective homomorphism
\[    i^\ast \maps \G_\phi   \to \G_\psi \]
given by the natural projection
\[   \G_\phi = \prod_{v \in V_\phi} P_v \times_{\Ad} G
\;\to \;\prod_{v \in V_\psi} P_v \times_{\Ad} G = \G_\psi .\]
The action of $\G_\phi$ on $\A_\phi$ is related to that of
$\G_\psi$ on $\A_\psi$ by
\[       i^\ast(gA) = i^\ast(g)i^\ast(A) \]
for all $g \in \A_\phi$, $A \in \A_\phi$.
It follows that $i \maps C(\A_\psi) \to C(\A_\phi)$ restricts to
a one-to-one algebra homomorphism
\[       i \maps C(\A_\psi/\G_\psi) \to C(\A_\phi/\G_\phi) ,\]
and $i \maps L^2(\A_\psi) \to L^2(\A_\phi)$ restricts to an
isometry
\[       i \maps L^2(\A_\psi/\G_\psi) \to L^2(\A_\phi/\G_\phi) .\]

In short, when the graph $\psi$ is included in $\phi$, we can
think of $L^2(\A_\psi/\G_\psi)$ as a subspace of $L^2(\A_\phi/\G_\phi)$.
The spin network states for $L^2(\A_\psi/\G_\psi)$ are then
automatically spin network states for $L^2(\A_\phi/\G_\phi)$:

\begin{lemma} \et \label{lem3} Suppose $\psi$ and $\phi$ are graphs,
$P_\psi \to V_\psi$ and $P_\phi \to V_\phi$ are
principal $G$-bundles, and $i \maps \psi \hookrightarrow \phi$ is an
inclusion such that $i^\ast P_\phi = P_\psi$.  Then the induced isometry
\[   i \maps L^2(\A_\psi/\G_\psi) \to L^2(\A_\phi/\G_\phi) \]
maps spin network states for the former space into
spin network states for the latter space.
\end{lemma}

Proof - We assume without loss of generality that
$V_\psi$ is a subset of $V_\phi$, $i \maps V_\psi \to V_\phi$
is given by $i(v) = v$, hence that bundle $P_\psi$ is
the restriction of $P_\phi$ to $V_\psi$.   Note that there is
a natural way to compose inclusions, and that any inclusion of
the above sort can be written as a product of a finite sequence of
inclusions, each of which is of one of the following four forms:

\begin{enumerate}

\item Adding a vertex:

\vskip1em
\vbox{
\begin{center}
\setlength{\unitlength}{0.00625in}%
\begin{picture}(120,36)(245,525)
\thicklines
\put(360,540){\circle*{10}}
\put(245,545){\line( 1, 0){ 45}}
\put(245,535){\line( 1, 0){ 45}}
\put(300,540){\line(-5, 3){ 25}}
\put(300,540){\line(-5,-3){ 25}}
\put(355,555){\makebox(0,0)[lb]{\raisebox{0pt}[0pt][0pt]{\twlrm $v$}}}
\end{picture}
\end{center}
}

More precisely, $V_\phi$ is the disjoint union of $V_\psi$
and $\{v\}$, and $E_\phi = E_\psi$.  The source and target
functions are the same for $\phi$ as for $\psi$, and the
inclusion $i \maps \psi \to \phi$ sends each edge $e$ of $\psi$
to the path $e$ in $\phi$.

\item  Adding an edge:
\vskip1em
\vbox{
\begin{center}
\setlength{\unitlength}{0.00625in}%
\begin{picture}(390,39)(75,525)
\thicklines
\put( 80,540){\circle*{10}}
\put(460,540){\circle*{10}}
\put(200,540){\circle*{10}}
\put(340,540){\circle*{10}}
\put(245,545){\line( 1, 0){ 45}}
\put(245,535){\line( 1, 0){ 45}}
\put(300,540){\line(-5, 3){ 25}}
\put(300,540){\line(-5,-3){ 25}}
\put(400,540){\vector( 1, 0){ 10}}
\put(340,540){\line( 1, 0){120}}
\put( 75,555){\makebox(0,0)[lb]{\raisebox{0pt}[0pt][0pt]{\twlrm $v_1$}}}
\put(195,555){\makebox(0,0)[lb]{\raisebox{0pt}[0pt][0pt]{\twlrm $v_2$}}}
\put(335,555){\makebox(0,0)[lb]{\raisebox{0pt}[0pt][0pt]{\twlrm $v_1$}}}
\put(455,555){\makebox(0,0)[lb]{\raisebox{0pt}[0pt][0pt]{\twlrm $v_2$}}}
\put(400,555){\makebox(0,0)[lb]{\raisebox{0pt}[0pt][0pt]{\twlrm $e$}}}
\end{picture}
\end{center}
}

Here $V_\phi = V_\psi$, for some $e \notin
E_\psi$ we have $E_\phi = E_\psi \cup \{e\}$,
and the source and target functions for $\phi$ agree with those of
$\psi$ on $E_\psi$, while
\[       s(e) = v_1, \qquad t(e) = v_2 \]
for some $v_1, v_2 \in V_\psi$.  The inclusion $i \maps
\psi \hookrightarrow \phi$ assigns to each edge $f$ of $\psi$
the path $f$ in $\phi$.

\item Subdividing an edge:
\vskip 1em
\vbox{

\begin{center}
\setlength{\unitlength}{0.00625in}%
\begin{picture}(390,39)(75,525)
\thicklines
\put( 80,540){\circle*{10}}
\put(200,540){\circle*{10}}
\put(340,540){\circle*{10}}
\put(400,540){\circle*{10}}
\put(460,540){\circle*{10}}
\put( 80,540){\line( 1, 0){120}}
\put(340,540){\line( 1, 0){120}}
\put(365,540){\vector( 1, 0){ 10}}
\put(430,540){\vector( 1, 0){  5}}
\put(135,540){\vector( 1, 0){ 20}}
\put(245,545){\line( 1, 0){ 45}}
\put(245,535){\line( 1, 0){ 45}}
\put(300,540){\line(-5, 3){ 25}}
\put(300,540){\line(-5,-3){ 25}}
\put( 60,555){\makebox(0,0)[lb]{\raisebox{0pt}[0pt][0pt]{\twlrm $s(e)$}}}
\put(195,555){\makebox(0,0)[lb]{\raisebox{0pt}[0pt][0pt]{\twlrm $ t(e)$}}}
\put(310,555){\makebox(0,0)[lb]{\raisebox{0pt}[0pt][0pt]{\twlrm $s(e)$}}}
\put(395,555){\makebox(0,0)[lb]{\raisebox{0pt}[0pt][0pt]{\twlrm $v$}}}
\put(455,555){\makebox(0,0)[lb]{\raisebox{0pt}[0pt][0pt]{\twlrm $t(e)$}}}
\put(135,555){\makebox(0,0)[lb]{\raisebox{0pt}[0pt][0pt]{\twlrm $e$}}}
\put(365,555){\makebox(0,0)[lb]{\raisebox{0pt}[0pt][0pt]{\twlrm $e_1$}}}
\put(425,555){\makebox(0,0)[lb]{\raisebox{0pt}[0pt][0pt]{\twlrm $e_2$}}}
\end{picture}
\end{center}
}
\vskip 1em

For some $v \notin V_\psi$ we have
$V_\phi = V_\psi \cup \{v\}$,
for some $e \in E_\psi$ and $e_1, e_2 \notin E_\psi$ we
have $E_\phi = (E_\psi - \{e\}) \cup \{e_1, e_2\}$,
and the source and target functions of $\phi$ agree with those of
$\psi$ on $E_\psi - \{e\}$, while
\[          s(e_1) = s(e), \qquad t(e_1) = v, \]
\[          s(e_2) = v, \qquad t(e_2) = t(e) .\]
The inclusion $i \maps \psi \hookrightarrow \phi$
assigns to each edge $f\ne e$ of $\psi$ the path $f$ in $\phi$,
and assigns to the edge $e$ the path $e_1 e_2$.

\item  Reversing the orientation of an edge:
\vskip 1em
\vbox{
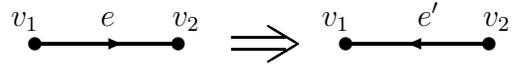
\begin{center}
\setlength{\unitlength}{0.00625in}%
\begin{picture}(390,39)(75,525)
\thicklines
\put( 80,540){\circle*{10}}
\put(460,540){\circle*{10}}
\put(200,540){\circle*{10}}
\put(340,540){\circle*{10}}
\put(245,545){\line( 1, 0){ 45}}
\put(245,535){\line( 1, 0){ 45}}
\put(300,540){\line(-5, 3){ 25}}
\put(300,540){\line(-5,-3){ 25}}
\put( 80,540){\line( 1, 0){120}}
\put(145,540){\vector( 1, 0){ 10}}
\put(340,540){\line( 1, 0){120}}
\put(400,540){\vector(-1, 0){ 10}}
\put( 60,555){\makebox(0,0)[lb]{\raisebox{0pt}[0pt][0pt]{\twlrm $v_1$}}}
\put(195,555){\makebox(0,0)[lb]{\raisebox{0pt}[0pt][0pt]{\twlrm $v_2$}}}
\put(320,555){\makebox(0,0)[lb]{\raisebox{0pt}[0pt][0pt]{\twlrm $v_1$}}}
\put(455,555){\makebox(0,0)[lb]{\raisebox{0pt}[0pt][0pt]{\twlrm $v_2$}}}
\put(400,555){\makebox(0,0)[lb]{\raisebox{0pt}[0pt][0pt]{\twlrm $e'$}}}
\put(135,555){\makebox(0,0)[lb]{\raisebox{0pt}[0pt][0pt]{\twlrm $e$}}}
\end{picture}
\end{center}
}

Here $V_\psi = V_\phi$, and
for some $e \in E_\psi$ and $e' \notin E_\psi$
we have $E_\phi = (E_\psi - \{e\}) \cup \{e'\}$.
The source and target functions for $\phi$ agree with those of
$\psi$ on $E_\psi - \{e\}$, while if
\[      s(e) = v_1, \qquad t(e) = v_2 \]
then
\[      s(e') = v_2 , \qquad t(e') = v_1 .\]
The inclusion $i \maps \psi \hookrightarrow \phi$ assigns to
each edge $f \ne e$ of $\psi$ the path $f$ in $\phi$, but
assigns to $e$ the path $e'^{-1}$.
\end{enumerate}

Thus, to show that $i$ maps each spin network state $\Psi_{\rho, \iota}
\in L^2(\A_\psi/\G_\psi)$ into a spin network state $\Psi_{\rho',
\iota'} \in L^2(\A_\phi/\G_\phi)$, it suffices to show this in each
of the four cases above.  Calculations give the following results for each
case:

\begin{enumerate}
\item  $i(\Psi_{\rho,\iota}) =
\Psi_{\rho', \iota'}$ is given as follows: $\rho' = \rho$, and
$\iota'_w = \iota_w$ for all vertices $w \in V_\psi$, while $\iota'_v =
1 \in \C$.  (Note that the $S(v)$ and $T(v)$ are the empty set,
and the empty tensor product of representations is defined to be
the trivial representation $\C$, so $\Inv_\phi(v,\rho') = \C$.)

\item  Here $\rho'_f = \rho_f$ for all edges $f \in E_\psi$ except
$e$, while $\rho_e = \C$.  Moreover, $\iota'_v = \iota_v$ for all
$v \in V_\psi$ except $v_1$ and $v_2$.  Given
\[    \iota_{v_1} \in  \Inv_\psi(v_1,\rho) =
 \Inv\left( \bigotimes_{f \in S_\psi(v)}
\rho_f  \otimes  \bigotimes_{f \in T_\psi(v)}
\rho_f^\ast \right), \]
then $\iota'_{v_1}$ is the vector in
\[      \Inv_\phi(v_1,\rho') =
 \Inv\left(\C \tensor \bigotimes_{f \in S_\psi(v)}
\rho_f  \otimes  \bigotimes_{f \in T_\psi(v)}
\rho_f^\ast \right) \]
corresponding to $\iota_{v_1}$ under the natural isomorphism
between these spaces.  The case of $v_2$ is analogous.

\item  Here $\rho'_f = \rho_f$ for all edges $f \in E_\psi$ except
$e_1$ and $e_2$, while $\rho'_{e_1} = \rho'_{e_2} = \rho_e$.
Moreover, $\iota'_w = \iota_w$ for all $w \in V_\psi$ except $v$, while
$\iota'_v$ is the vector in $ \Inv(\rho_e \tensor
\rho_e^\ast)$ corresponding to the identity intertwining operator
$1 \maps \rho_e \to \rho_e$.

\item  Here $\rho'_f = \rho_f$ for all edges $f \in E_\psi$ except
$e'$, while $\rho'_{e'} = \rho_e^\ast$.  Moreover, $\iota'_v = \iota_v$
for all $v \in V_{\psi}$ except $v_1$ and $v_2$.
Given
\ban   \iota_{v_1} \in \Inv_\psi(v_1,\rho) &=&
\Inv\left( \bigotimes_{f \in S_\psi(v_1)}
\rho_f \, \tensor  \bigotimes_{f \in T_\psi(v_1)}
\rho_f^\ast \right) \\
&\cong& \Inv\left(\rho_e \tensor \bigotimes_{f \in S_\psi(v_1), f \ne e}
\rho_f \, \tensor  \bigotimes_{f \in T_\psi(v_1)}
\rho_f^\ast \right), \ean
then $\iota'_{v_1}$ is the vector in
\ban     \Inv_\phi(v_1, \rho') &=&
\Inv \left( \bigotimes_{f \in S_\phi(v_1)}
\rho_f \, \tensor  \bigotimes_{f \in T_\phi(v_1)}
\rho_f^\ast \right) \\
&\cong&  \Inv\left(\rho_e \tensor \bigotimes_{f \in S_\phi(v_1)}
\rho_f \, \tensor  \bigotimes_{f \in T_\phi(v_1), f \ne e'}
\rho_f^\ast \right)
\ean
corresponding to $\iota_{v_1}$ under the natural isomorphism between
these spaces.  The case of $v_2$ is analogous.  \qed
\end{enumerate}

The proof of Lemma \ref{lem3} not only shows that $i$ maps
each spin network state $\Psi_{\iota,\rho} \in L^2(\A_\psi/\G_\psi)$
into a spin network state $\Psi_{\iota',\rho'} \in
L^2(\A_\phi/\G_\phi)$; it also gives an algorithm for computing
$\iota', \rho'$ from $\iota, \rho$.  This is likely to be useful in
applications.  It also yields:

\begin{lemma} \label{lem4} \et  Given the hypothesis of
Lemma \ref{lem3}, suppose $\Psi \in L^2(\A_\psi/\G_\psi)$ is such that
$i(\Psi)$ is a spin network state in $L^2(\A_\phi/\G_\phi)$.  Then
$\Psi$ is a spin network state.  \end{lemma}

Proof - Given any spin network state $\Psi_{\rho,\iota} \in
L^2(\A_\psi/\G_\psi)$, write
\[       i(\Psi_{\rho,\iota}) = \Psi_{\rho',\iota'}  .\]
By Lemma \ref{lem2} we can write
\[     \Psi = \sum_{\rho} \sum_\iota c_{\rho,\iota} \Psi_{\rho,\iota} \]
where $\rho$ ranges over $\Lambda^E$ and for each
$\rho$, $\iota$ ranges over an orthonormal basis of $\Inv_\psi(v,\rho)$.
Then
\[ i(\Psi) = \sum_{\rho} \sum_{\iota} c_{\rho,\iota} \Psi_{\rho',\iota'} .\]
Note from the proof of Lemma \ref{lem3} that $\rho'$ depends only on
$\rho$, not $\iota$, and the function $\rho \mapsto \rho'$ is
one-to-one.  Furthermore, if $i(\Psi)$ is spin network state all the
summands in the above equation must vanish except those involving
a particular choice of $\rho'$.  Thus for some particular choice of
$\rho$,
\[    i(\Psi) = \sum_{\iota} c_{\rho, \iota} \Psi_{\rho',\iota'} .\]
Now the requirement that $i(\Psi)$ be a spin network state implies
that this vector, which lies in
\[      \bigotimes_{v \in V_\phi} \Inv_\phi(v,\rho') ,\]
is a tensor product of vectors in the factors.   It follows that
\[    \Psi = \sum_{\iota} c_{\rho, \iota} \Psi_{\rho,\iota} \]
and that this vector, which lies in
\[      \bigotimes_{v \in V_\phi} \Inv_\phi(v,\rho') ,\]
is a tensor product of vectors in the factors.  Thus $\Psi$ is a spin
network state.  \qed

\section{The Loop Representation}

To apply the result of the previous section to gauge theory on a
manifold, we need to recall some facts about the loop
representation.   All the material in this section can be found in
existing mathematically rigorous work on the loop representation
\cite{AI,AL,ALMMT,Baez,Baez2,Baez3,L}

Let $M$ be a real-analytic manifold and let $P$ be a smooth
principal $G$-bundle over $M$, with $G$ a compact connected Lie group.
Let $\A$ be the space of smooth connections on $P$ and $\G$ the
group of smooth gauge transformations.  By a {\sl path} in $M$ we will
always mean a piecewise analytic path.  Given a path $\gamma$ in $M$,
let $\A_\gamma$ denote the space of smooth maps $F\maps P_{\gamma(a)}
\to P_{\gamma(b)}$ that are compatible with the right action of $G$ on
$P$:
\[            F(xg) = F(x)g .\]
Note that for any connection $A \in \A$, the parallel transport
map
\[        T \exp\int_\gamma A: P_{\gamma(a)} \to P_{\gamma(b)} \]
lies in $\A_\gamma$.  Of course, if we fix a
trivialization of $P$ at the endpoints of $\gamma$,
we can identify $\A_\gamma$ with the group $G$.

Let the algebra $\Fun_0(\A)$ of {\it cylinder functions}
be the algebra of functions on $\A$ generated by
those of the form
\[            F(T \exp \int_\gamma A)  \]
where $F$ is a continuous function on $\A_\gamma$.  Let $\Fun(\A)$
denote the completion of $\Fun_0(\A)$ in the sup norm:
\[          \|f \|_\infty = \sup_{A \in \A} |f(A)|. \]
Equipped with this norm, $\Fun(\A)$ is a commutative C*-algebra.
A {\sl generalized measure} on $\A$ is defined to be a continuous linear
functional $\nu \maps \Fun(\A) \to \C$.

We say that the generalized measure $\nu$ on $\A$ is {\sl strictly
positive} if $\nu(f) > 0$ for all nonzero $f \ge 0$ in
$\Fun(\A)$.   The group $\G$ acts as gauge transformations on $\A$,
and as automorphisms of $\Fun(\A)$ by
\[       gf(A) = f(g^{-1}A)  ,\]
where $A \in \A$.  We say that $\nu$ is {\sl
gauge-invariant} if for all $g \in \G$ and $f \in \Fun(\A)$ we
have $\nu(gf) = \nu(f)$.   In the next section we focus on a
particular gauge-invariant, strictly positive generalized measure
on $\A$,  the `uniform' generalized measure.  This serves as a
kind of substitute for the purely formal `Lebesgue measure' on
$\A$, but it is constructed using Haar measure on $G$ rather the
structure of $\A$ as an affine space.

In fact, any gauge-invariant, strictly positive generalized
measure $\nu$ on $\A$ allows us to define analogues of the space
of $L^2$ functions on $\A$ and $\A/\G$, as follows.   First, we
define the Hilbert space $L^2(\A,\nu)$ to be the completion of
$\Fun(\A)$ in the norm
\[         \|f\|_2 = \nu(|f|^2)^{1/2} .\]
Then, let $\Fun(\A/\G)$ to be the subalgebra of gauge-invariant
functions in $\Fun(\A)$.  These functions can also be regarded as
continuous functions $\A/\G$ with its quotient topology.
We define a {\sl generalized
measure} on $\A/\G$ to be a continuous linear functional from
$\Fun(\A/\G)$ to $\C$.  Any generalized measure $\nu$ on
$\A$ restricts to a generalized measure on $\A/\G$.
This restriction process defines a one-to-one
correspondence between gauge-invariant generalized measures on
$\A$ and generalized measures on $\A/\G$.   For example, the
uniform generalized measure on $\A$ corresponds to a measure on
$\A/\G$ called the `Ashtekar-Lewandowski' generalized measure
\cite{AL}.

Finally, given a gauge-invariant, strictly positive generalized
measure $\nu$ on $\A$,  define $L^2(\A/\G,\nu)$ to be the
completion of $\Fun(\A/\G)$ in the above norm $\|\cdot\|_2$.   It
turns out that the representation of $\G$ on $\Fun(\A)$ extends uniquely
to a unitary representation of $\G$ on $L^2(\A,\nu)$, and that
$L^2(\A/\G,\nu)$ is naturally isomorphic as a Hilbert space to
the subspace of $\G$-invariant elements of $L^2(\A)$.  Moreover,
the algebra $\Fun_0(\A/\G)$ of gauge-invariant cylinder functions
on $\A$ is dense in $\Fun(\A/\G)$, hence in $L^2(\A/\G,\nu)$.

\section{Spin Network States}

While none of the results on the loop representation in Section 3
explicitly mention graphs, the proofs of some, and the actual
construction of interesting generalized measures on $\A$, turn
out to be closely related to gauge theory on graphs
\cite{ALMMT,Baez,Baez2,Baez3,L}. In this section we recall this
relationship and use it to prove the main result about spin
network states.

There is an equivalence relation on paths $\gamma \maps [0,1] \to M$
that are embeddings when restricted to $(0,1)$, namely, $\gamma_1 \sim
\gamma_2$ if $\gamma_1$ is obtained from $\gamma_2$ by an
orientation-preserving continuous reparametrization with continuous
inverse.  We call an equivalence class $e = [\gamma]$ of such paths an
{\it embedded edge} in $M$.  Note that the endpoints $\gamma(0)$ and
$\gamma(1)$ are independent of a choice of representative $\gamma$ for
$e$, as is the set $\gamma[0,1] \subseteq M$.  We write these as $e(0)$,
$e(1)$, and $e[0,1]$, respectively.  We define an {\it embedded graph}
$\phi$ to be a finite collection $e_i$ of embedded edges such that for
all $i \ne j$, $e_i[0,1]$ and $e_j[0,1]$ intersect, if at all, only at
their endpoints.  We call $e_i$ {\it edges} of $\phi$, and call the
points $e_i(0),e_i(1)$ the {\it vertices} of $\phi$.  Somewhat redundantly,
we write $E_\phi$ for the set of edges of $\phi$, and $V_\phi$ for the
set of vertices.

Note that any graph $\phi$ embedded in $M$ determines a graph in
the sense of Section 2 ---
which by abuse of notation we also call $\phi$ ---
having edges $E_\phi$, vertices $V_\phi$, and
\[       s(e_i) = e_i(0) , \qquad
        t(e_i) = e_i(1) .\]
(Sometimes we will call graphs in the sense of Section 2 {\it
abstract} graphs, to distinguish them from embedded graphs.)
Moreover, if we restrict the bundle $P$ to the vertices of $\phi$,
we obtain a principal $G$-bundle $P_\phi$ over $V_\phi$.
We may thus define the space $\A_\phi$ of connections on $\phi$,
the group $\G_\phi$ of gauge transformations, and so on, as in Section 2.

The uniform generalized measure $\mu$ on $\A$ can be efficiently described
using embedded graphs, as follows.  There is an onto map
\[          p_\phi \maps \A \to \A_\phi  \]
given by
\[          (p_\phi(A))_e =   T \exp \int_\gamma A  ,\]
where $\gamma$ is any representative of the embedded edge $e$.
This map allows us to identify
$C(\A_\phi)$ with a subalgebra of the algebra $\Fun_0(\A)$ of cylinder
functions, and $\Fun_0(\A)$ is the union of the
algebras $C(\A_\phi)$ as $\phi$ ranges over all graphs embedded in $M$.
The uniform generalized measure $\mu$ on $\A$ is then uniquely
characterized by the property that for any embedded graph $\phi$, and
any $f \in C(\A_\phi)$,
\[         \mu(f) = \int_{\A_\phi} f \mu_\phi ,\]
where $\mu_\phi$ is the measure on $\A_\phi$ introduced in
Section 2 (essentially a product of copies of normalized Haar measure on
$G$).

As in the previous section
we define $L^2(\A)$ to be the completion of $\Fun(\A)$ in
the norm
\[        \|f\|_2 = \mu(|f|^2)^{1/2}  , \]
and $L^2(\A/\G)$ to be the completion of $\Fun(\A/\G)$
in the same norm.  By the defining property of $\mu$, the inclusion
\[       C(\A_\phi) \subset \Fun(\A_\phi)  \]
extends uniquely to an isometry
\[       L^2(\A_\phi) \to L^2(\A) \]
which in turn restricts to an isometry
\[       L^2(\A_\phi/\G_\phi) \to L^2(\A/\G) \]
Thus we can think of $L^2(\A_\phi)$ as a closed subspace of
$L^2(\A)$,  and $L^2(\A_\phi/\G_\phi)$ as a closed subspace of
$L^2(\A/\G)$.   By results recalled in Section 3,  the union of
the subspaces $L^2(\A_\phi)$ is dense in $L^2(\A)$, and similarly
the union of the subspaces $L^2(\A_\phi/\G_\phi)$ is dense in
$L^2(\A/\G)$.

It is by this method that we obtain spin network states spanning
$L^2(\A/\G)$.  Given any graph $\phi$ embedded in $M$, any labelling
$\rho$ of the edges $e$ of $\phi$ by irreducible representations $\rho_e
\in \Lambda$ of $G$, and any labelling of the vertices $v$ of $\phi$ by
vectors in $\Inv(v,\rho)$, we write $\Psi_{\phi,\rho,\iota}$ for the
spin network state $\Psi_{\rho,\iota}$ thought of as a vector in
$L^2(\A/\G)$.

\begin{theorem} \et   The set of all
vectors of the form $\Psi_{\phi,\rho,\iota}$ spans
$L^2(\A/\G)$.  \end{theorem}

Proof - The union of the subspaces $L^2(\A_\phi/\G_\phi)$ is dense in
$L^2(\A/\G)$, and for each embedded graph $\phi$
the vectors $\Psi_{\rho,\iota}$ span
$L^2(\A_\phi/\G_\phi)$, by Lemma \ref{lem2.5}  \qed

We call the vectors $\Psi_{\phi,\rho,\iota}$ {\it spin network states}.
Note that this concept is unambiguous, in the sense that the question of
whether a given vector in $L^2(\A/\G)$ is a spin network state can be
answered irrespective of a choice of embedded graph:

\begin{theorem}\et Given
\[  \Psi_{\phi,\rho,\iota} \in L^2(\A_\phi/\G_\phi) \subseteq L^2(\A/\G) \]
for some embedded graph $\phi$, if
\[  \Psi_{\phi,\rho,\iota} \in L^2(\A_{\phi'}/\G_{\phi'}) \subseteq L^2(\A/\G)
\]
for some other embedded graph $\phi'$, then
\[  \Psi_{\phi,\rho,\iota} = \Psi_{\phi',\rho',\iota'} \]
for some labellings $\rho'$ and $\iota'$. \end{theorem}

Proof - First note \cite{Baez} that given two embedded graphs, there is
always a third embedded graph including both, so that it suffices to
consider the cases where $\phi \hookrightarrow \phi'$ or
$\phi' \hookrightarrow \phi$.   In the former case, we obtain
$\Psi_{\phi,\rho,\iota} = \Psi_{\phi',\rho',\iota'}$ using Lemma 3.
In the latter case, the result follows using Lemma 4.  \qed

\section{Conclusions}

One aim of this work is to provide tools for
work on the loop representation of quantum gravity and other
diffeomorphism-invariant gauge theories.   In the loop
representation of quantum gravity \cite{RS}, it is typical to
proceed towards the description of physical states in three
stages.  The first two stages have been formalized in a
mathematically rigorous way, but until the crucial third stage
has been dealt with rigorously, the success of the whole program
is an open question.  In particular, it is quite possible that
the work done so far will need refinement and revision to provide
a sufficient platform for the third stage.

In the first stage, {\it kinematical states} are taken to be
generalized measures on $\A/\G$, where $\A$ is the space of
$\SU(2)$ connections on a bundle isomorphic to the spin bundle of
the (real-analytic, oriented) 3-manifold $M$ representing
`space' in the theory.  Such generalized measures can be
characterized in  terms of embedded graphs \cite{Baez}.  Briefly,
they are in one-to-one correspondence with `consistent' uniformly
bounded families of measures on the spaces $\A_\phi/\G_\phi$ for
all graphs $\phi$ embedded in $M$.  Here {\it consistency} means that
when the embedded graph $\psi$ is included in $\phi$, the measure
on $\A_\phi/\G_\phi$ must push forward to the measure on
$\A_\psi/\G_\psi$ under the induced map from $\A_\phi/\G_\phi$ to
$\A_\psi/\G_\psi$.

In the second stage, {\it diffeomorphism-invariant states} are taken to
be generalized measures on $\A/\G$ that are invariant under the action
of $\Diff_0(M)$, the identity component of the group of real-analytic
diffeomorphisms of $M$.  (It is worth noting that analyticity plays a
technical role here and a purely $C^\infty$ version of the theory
would be preferable in some ways.)  Diffeomorphism-invariant states have
also been characterized in terms of embedded graphs \cite{Baez}.
Unfortunately, while this characterization allows the explicit
construction of many diffeomorphism-invariant states, of which the
Ashtekar-Lewandowski generalized measure is the simplest, it does not
give a concrete recipe for constructing `all' diffeomorphism-invariant
states, or even a dense set.  In particular, the `loop states', so
important in the heuristic work of Rovelli and Smolin \cite{RS}, remain
mysterious from this viewpoint.

Finally, one hopes that {\it physical states} are
diffeomorphism-invariant states that satisfy a certain constraint, the
Hamiltonian constraint.  Formulating this constraint rigorously is a key
technical problem in the loop representation of quantum gravity, much
studied \cite{BP} but still insufficiently understood.  One key aspect,
the interplay between $\SU(2)$ and $\SL(2,\C)$ connections which is so
important in the theory, has recently been clarified using embedded
graph techniques \cite{ALMMT}.  But one would also like to make precise
various arguments such as Rovelli and Smolin's argument that the `loop
states' satisfy the Hamiltonian constraint.  In order to do this, it is
important to understand the diffeomorphism-invariant states as
explicitly as possible.

In this direction, recent work by Ashtekar and collaborators
\cite{Ashtekar3,AL2} has given a rigorous construction of the
Rovelli-Smolin `loop states'.  The construction is applicable to
any compact connected gauge group $G$, and it produces a
diffeomorphism-invariant generalized measure $\nu$ on $\A/\G$
from an isotopy equivalence class of knots $K$ and an irreducible
representation $\rho$ of $G$, as follows.   For the present
purposes, we define a {\it knot} to be an equivalence class of
analytically embedded circles in $M$, two embeddings being
equivalent if they differ by an orientation-preserving continuous
reparametrization with continuous inverse.   Also, we define two
knots to be {\sl isotopic} if one can be obtained from the other
by the action of $\Diff_0(M)$.   Now, given an isotopy equivalence
class of knots $K$, choose for each knot $k \in K$ an analytically embedded
circle $\gamma \maps S^1 \to M$ representing $k$.  Note that
$\gamma$, regarded as a map from $[0,1]$ to $M$ with
$\gamma(0) = \gamma(1)$, defines an embedded edge $e$ in
$M$.  Associated to $e$ there is an embedded graph $\phi$
having $e$ as its only edge and $v = e(0) = e(1)$ as its
only vertex.   If we label the edge $e$ with the
representation $\rho \in \Lambda$ and label the vertex $v$ with
the identity operator (as an intertwining operator from $\rho$ to
itself), we obtain a spin network state $\psi_k \in L^2(\A/\G)$.
Next, consider the formal sum over all knots $k$ in the isotopy
class $K$,
\[         \nu = \sum_{k \in K} \psi_k  .\]
This sum does not converge in
$L^2(\A/\G)$ --- indeed, it is an uncountable sum --- but one can
show that for any function $f \in \Fun(\A/\G)$, the sum
\[    \nu(f) =  \sum_{k \in K} \langle \psi_k,f \rangle \]
does converge, where the inner product is that of $L^2(\A/\G)$.
In fact, $\nu$ defines a generalized measure on $\A/\G$.
By the nature of the construction it is clear that $\nu$ is
$\Diff_0(M)$-invariant.  We call $\nu$ the {\it loop state}
associated to the knot class $K$ and the representation $\rho$.

The framework of spin network states appears to allow an
interesting generalization of the above construction.   Namely,
one should be able to construct diffeomorph-\break ism-invariant
generalized measures on $\A/\G$ from isotopy classes of spin
networks.  The key idea is to treat a knot labelled by a group
representation as a very special case of a spin network, in a
manner that our exposition above should make clear.  This idea,
currently under investigation by Ashtekar and the author, might
make possible the sort of explicit description of `all'
diffeomorphism-invariant states that one would like for
rigorous work on the Hamiltonian constraint.

Another aim of this work is to clarify the relationship between
the loop representation of quantum gravity and a body of recent
work on topological quantum field theories \cite{CKY,KL,RT,TV}.
As noted in Section 1, spin networks arise naturally in
the study of the category of representations of a group or
quantum group.  More generally, we may define them for any
category with the appropriate formal properties.  It is a
striking fact that the most efficient construction of many
topological quantum field theories involves category theory and
the use of spin networks.  For example, Euclidean 3-dimensional
quantum gravity with nonzero cosmological constant can be
identified with the Turaev-Viro theory \cite{TV}, and
the latter is a topological quantum field theory that is most
easily constructed using $SU_q(2)$ spin networks. It is unclear
whether 4-dimensional quantum gravity is a topological quantum
field theory (or some generalization thereof), but the present
work at least begins to make precise the role of spin networks in the loop
representation of 4-dimensional quantum gravity.   In what
follows we briefly comment on the category-theoretic significance
of our results.

Associated to any abstract graph $\phi$ in the sense of Section 2 there
is a category $\c_\phi$, or more precisely, a groupoid (a category in which
all the morphisms are invertible).  This is the free groupoid on
the objects $V_\phi$ and morphisms $E_\phi$.  If we fix a trivial
$G$-bundle $P$ over $V_\phi$, the connections $A \in \A_\phi$ are
precisely the functors from $\c_\phi$ to $G$, where we regard the
compact connected Lie group $G$ as a groupoid with one object.
Similarly, the gauge transformations $g \in \G_\phi$ are
precisely the natural transformations between such functors.  As
we have seen, the set $\A_\phi/\G_\phi$ of `functors modulo
natural transformations' inherits the structure of a measure
space from $G$, and Lemma \ref{lem2.5} gives an explicit
description of $L^2(\A_\phi/\G_\phi)$ in terms of the category of
finite-dimensional unitary representations of $G$.

Similarly, given a real-analytic manifold $M$ and a smooth principal
$G$-bundle $P$ over $M$, we may define the {\it holonomy groupoid} $\c$
to have as objects points of $M$ and as morphisms equivalence classes of
piecewise analytic paths in $M$, where two paths $\gamma,\gamma'$ are
regarded as equivalent if
\[ T\exp \int_\gamma A = T\exp \int_{\gamma'} A \]
for all connections $A$ on $P$.  This has as a
subgroupoid the `holonomy loop group' of Ashtekar and Lewandowski
\cite{AL}.  If we fix a trivialization of $P_x$ for all $x \in M$, any
connection on $P$ determines a functor from $\c$ to $G$, while
conversely any such functor can be thought of as a `generalized
connection' \cite{Baez2,Baez3}.  Similarly, any gauge transformation
determines a natural transformation between such functors, and any
natural transformation between such functors can be thought of as a
`generalized gauge transformation'.

The relation between gauge theory on graphs and gauge theory on
manifolds then turns upon the fact that for any graph $\phi$
embedded in $M$ we obtain a subcategory of $\c$ isomorphic to
$\c_\phi$.  Moreover, an inclusion $i \maps \psi \hookrightarrow \phi$
induces a functor $i_\ast \maps \c_\psi \hookrightarrow \c_\phi$,
and the holonomy groupoid $\c$ is the colimit of the groupoids
$\c_\phi$ as $\phi$ ranges over all embedded graphs in $M$.
This explains the importance of `projective limit' techniques
in studying generalized measures on the space of connections and
the space of connections modulo gauge transformations \cite{AL2}.
In particular, it is this that lets us obtain a spanning set of
spin network states for $L^2(\A/\G)$ from the spin network states
for $L^2(\A_\phi/\G_\phi)$ as $\phi$ ranges over all graphs
embedded in $M$.

Finally, it is interesting to note that the holonomy groupoid
$\c$ has as a quotient the fundamental groupoid of $M$, in which
morphisms are given by homotopy equivalence classes of paths.
A functor from $\c$ to $G$ that factors through the fundamental
groupoid is just a flat connection on $P$.  In certain cases
there is a natural measure on the space of $\A_0/\G$ of flat connections
modulo gauge transformations, and then the space $L^2(\A_0/\G)$
is the Hilbert space for a theory closely related to quantum
gravity, namely $BF$ theory \cite{Baez3}.

\section*{Acknowledgements}

The author would like to thank Abhay Ashtekar, Carlo Rovelli and
Lee Smolin for a very inspiring conversation on spin networks in
quantum gravity, and also James Dolan for many fascinating
conversations about category theory.


\begin{thebibliography}{10}
\bibitem{Ashtekar} A.\ Ashtekar,
Mathematical problems of non-perturbative quantum general relativity, to
appear in {\sl Proceedings of the 1992 Les Houches Summer School on
Gravitation and Quantization,} ed.\ B.\ Julia, North-Holland, Amsterdam,
1993.

\bibitem{Ashtekar3} A.\ Ashtekar, Recent mathematical developments in
quantum general relativity, to appear in the Proceedings of the Seventh
Marcel Grossman Meeting on General Relativity.

\bibitem{AI} A.\ Ashtekar and C.\ J.\ Isham, Representations of the holonomy
algebra of gravity and non-abelian gauge theories, {\sl Class.\
Quan.\ Grav.\ }{\bf 9} (1992), 1069-1100.

\bibitem{AL} A.\ Ashtekar and J.\ Lewandowski, Representation
theory of analytic holonomy C*-algebras, in {\sl Knots and
Quantum Gravity,} ed.\ J.\ Baez, Oxford, Oxford U.\ Press, 1994.

\bibitem{AL2} A.\ Ashtekar and J.\ Lewandowski, Projective techniques
and functional integration for gauge theories, to appear.

\bibitem{ALMMT} A.\ Ashtekar, J.\ Lewandowski, D.\ Marolf,
J.\ Mour\~ao and T.\ Thiemann, Coherent state transforms for spaces of
connections, to appear.

\bibitem{Baez} J.\ Baez, Diffeomorphism-invariant generalized
measures on the space of connections modulo gauge
transformations, in {\it Proceedings of the Conference on Quantum Topology,}
ed.\ D.\ Yetter, World Scientific, Singapore, 1994.

\bibitem{Baez2} J.\ Baez, Generalized measures in gauge theory, {\sl Lett.\
Math.\ Phys.\ }{\bf 31} (1994), 213-223.

\bibitem{Baez3} J.\ Baez, Knots and quantum gravity: progress and
prospects, to appear in the Proceedings of the Seventh Marcel Grossman
Meeting on General Relativity, preprint available as gr-qc/9410018.

\bibitem{BP} B.\ Br\"{u}gmann and J.\ Pullin,
On the constraints of quantum gravity in the loop representation,
{\sl Nucl.\ Phys.\ }{\bf B390} (1993), 399-438

\bibitem{CKY} L.\ Crane, L.\ Kauffman and D.\ Yetter, State-sum invariants of
4-manifolds, I, preprint available as hep-th/9409167.

\bibitem{KL} L.\ Kauffman and S.\ Lins, {\sl Temperley-Lieb Recoupling Theory
and Invariants of 3-Manifolds}, Princeton U.\ Press, Princeton, 1994.

\bibitem{L} J.\ Lewandowski, Topological measure and
graph-differential geometry on the quotient space of connections,
International Journal of Theoretical Physics, {\bf 3} (1994) 207-211.

\bibitem{Penrose}
R.\ Penrose, Angular momentum; an approach to combinatorial space
time, in {\sl Quantum Theory and Beyond,}  ed.\ T.\ Bastin,
Cambridge University Press, Cambridge, 1971.

\bibitem{RT} N.\ Reshetikhin and V.\ Turaev, Ribbon
graphs and their invariants derived from quantum groups,
{\sl Comm.\ Math.\ Phys.\ }{\bf 127} (1990), 1-26.

Invariants of 3-manifolds via link polynomials and quantum groups, {\sl
Invent.\ Math.\ }{\bf 103} (1991), 547-597.

\bibitem{RS} C.\ Rovelli and L.\ Smolin, Loop representation for
quantum general relativity, {\sl Nucl.\ Phys.\ }{\bf B331} (1990),
80-152.

\bibitem{RS2} C.\ Rovelli, `In search of the topological Feynman rules
for quantum gravity,' lecture at Second Pennsylvania State Conference:
Quantum Geometry, Sept.\ 2, 1994.

C.\ Rovelli and S.\ Smolin, Discreteness of area and volume in quantum
gravity, preprint available as gr-qc/9411005.

\bibitem{TV} V.\ Turaev and O.\ Viro, State sum invariants of 3-manifolds and
quantum $6j$ symbols, {\sl Topology} {\bf 31} (1992), 865-902.

\end{thebibliography}
\end{document}